\documentclass
[preprint,prb,a4paper,showpacs,amsfonts,amssymb,final,notitlepage,tightenlines,11pt]{revtex4}%
\usepackage{amsfonts}
\usepackage{amsmath}
\usepackage{amssymb}
\usepackage{graphicx}%
\newtheorem{theorem}{Theorem}
\newtheorem{acknowledgement}[theorem]{Acknowledgement}

\begin{document}
\author{M. El Massalami,}
\affiliation{Instituto de Fisica, Universidade Federal do Rio de Janeiro, CxP 68528,
21945-970 Rio de Janeiro, Brazil}
\author{H. Takeya, K. Hirata,}
\affiliation{National Institute for Materials Science,1-2-1
Sengen,Tsukuba,Ibaraki,305-0047, Japan,}
\author{M. Amara, R.-M. Galera and D. Schmitt}
\affiliation{Labo. Magnetisme Louis Neel - CNRS, BP 166, 38042 Grenoble Cedex 9, France.}
\title{Magnetic Phase Diagram of GdNi$_{2}$B$_{2}$C: Two-ion Magnetoelasticity and
Anisotropic Exchange Couplings. }
\date{\today}

\begin{abstract}
Extensive magnetization and magnetostriction measurements were carried out on
a single crystal of GdNi$_{2}$B$_{2}$C\ along the main tetragonal axes. Within
the paramagnetic phase, the magnetic and strain susceptibilities revealed a
weak anisotropy in the exchange couplings and two-ion tetragonal-preserving
$\alpha$-strain modes. Within the ordered phase, magnetization and
magnetostriction revealed a relatively strong orthorhombic distortion mode and
rich field-temperature phase diagrams. For \textit{H}//(100) phase diagram,
three field-induced transformations were observed, namely, at: $H_{D}(T)$,
related to the domain alignment; $H_{R}(T)$, associated with reorientation of
the moment towards the c-axis; and $H_{S}(T)$, defining the saturation process
wherein the exchange field is completely counterbalanced. On the other hand,
For \textit{H}//(001) phase diagram, only two field-induced transformations
were observed, namely at: $H_{R}(T)$ and $H_{S}(T)$. For both phase diagrams,
$H_{S}(T)$ follows the relation $H_{S}\left[  1-\left(  T/T_{N}\right)
^{2}\right]  ^{\frac{1}{2}}$ kOe with $H_{S}(T\rightarrow0)$=128.5(5) kOe and
$T_{N}(H=0)$=19.5 K. In contrast, the thermal evolution of $H_{R}(T)$ along
the c-axis (much simpler than along the a-axis) follows the relation
$H_{R}\left[  1-T/T_{R}\right]  ^{\frac{1}{3}}$ kOe where $H_{R}%
(T\rightarrow0)$=33.5(5) kOe and $T_{R}(H=0)$=13.5 K. It is emphasized that
the magnetoelastic interaction and the anisotropic exchange coupling are
important perturbations and therefore should be explicitly considered if a
complete analysis of the magnetic properties of the borocarbides is desired.

\end{abstract}
\pacs{75.30.Kz,75.30.Cr,75.80.+q}
\maketitle

\section{Introduction}

The field-temperature ($H-T$) magnetic phase diagrams of the intermetallic
borocarbides $R$Ni$_{2}$B$_{2}$C ($R$=magnetic rare earth) manifest wide
varieties of zero-field magnetic structures such as Neel-type, helimagnetic,
amplitude-modulated, and squared-up states.$^{1-8}$ Most of these states, in
particular the modulated ones, are unstable against field and temperature
variation, leading to a cascade of transformations
\cite{Er-HT-diagram,Ho-HT-Diagram,Dy-HT-diagram} similar to the ones observed
in the elemental rare-earths. It is the general opinion that the remarkable
features observed in these $H-T$ phase diagrams are governed by the combined
influence of exchange couplings\ and crystalline electric field (CEF)
interactions \cite{Theory-exchange-CEF-Phase-diagram}: such an approach
reproduced successfully\ the gross features of the magnetic $H-T$ phase
diagrams of HoNi$_{2}$B$_{2}$C. Nevertheless, various experimental
observations suggest that two additional interactions, namely, anisotropic
exchange interaction and magnetoelastic coupling, are necessary ingredients
for the understanding of the behavior of the magnetic borocarbides.
Magnetoelastic interactions are spectacularly manifested in the onset of a
tetragonal-to-orthorhombic distortion at $T_{N}$ for $R$= Er, Ho, Dy, Tb
compounds
\cite{Er-magnetostriction,Ho-distortion,Dy-magnetostriction,Tb-dichroism-HRMXRD}%
, while the anisotropic exchange interactions were reported to be necessary
for the description of the low-temperature magnetic properties of ErNi$_{2}%
$B$_{2}$C (Ref.\cite{Er-magnetostriction}) and GdNi$_{2}$B$_{2}$C
(Ref.\cite{Gd-single-crystal-Tdynamics}). Obviously, the anisotropic exchange
interactions and (two-ion) magnetoelastic couplings can be conveniently
investigated in GdNi$_{2}$B$_{2}$C wherein CEF anisotropy is negligible and
the de Gennes factor is the strongest.

This work reports a study of the magnetic and magnetoelastic properties of
single crystal GdNi$_{2}$B$_{2}$C. The $H-T$ magnetic phase diagrams for
fields along the a- and c-axis were determined. Moreover, within the
paramagnetic phase, the parastriction and paramagnetic measurements revealed
the presence of weakly anisotropic exchange interaction as well as two-ion
magnetoelastic couplings while within the ordered state these perturbations
were observed to be strong enough to induce a noticeable modification in the
nuclear and magnetic structures.

GdNi$_{2}$B$_{2}$C crystallizes in a body-centered tetragonal structure (space
group I4/mmm, point symmetry of the Gd site is 4/mmm).\cite{R1221-structure}
The zero-field magnetic state immediately below \textit{T}$_{N}=19.5$ K is a
transverse, incommensurate, sine-modulated structure\ (moments along b-axis)
with a wave vector $\overrightarrow{k_{a}}$ that decreases linearly from
0.551$a^{\ast}$ at $T_{N}$ to 0.550$a^{\ast}$ at $T_{R}\approx$13.5 K.
\cite{Gd-XRES,Gd-single-crystal-Tdynamics,Gd-poly-high-field,Gd-chemical-composition,Gd-MES}
At $T_{R}$, a moment reorientation sets-in leading to an additional modulated
component transversely polarized along the c-axis.\cite{Gd-XRES,Gd-MES} On
further temperature decrease, $\overrightarrow{k_{a}}$ increases monotonically
reaching 0.553$a^{\ast}$ at 3.5 K.

\section{Theoretical Backgrounds}

{}The $^{7/2}S$-character of GdNi$_{2}$B$_{2}$C leads to a greater
simplification of its magnetic Hamiltonian which is considered to consist
mainly of an exchange, a Zeeman, a two-ion magnetoelastic, and an elastic
term. Within the mean-field approximation (MFA) the tetragonal-invariant
bilinear exchange interaction can be written as a sum of two
terms:\cite{Gignoux-Schmitt-1990}
\begin{align}
\mathcal{H}_{ex}^{\alpha_{1}} &  =-(g_{_{J}}\mu_{B})^{2}n^{\alpha_{1}%
}<J>J,\nonumber\\
\mathcal{H}_{ex}^{\alpha_{2}} &  =-(g_{_{J}}\mu_{B})^{2}n^{\alpha_{2}}\left(
2<J_{z}>J_{z}-<J_{x}>J_{x}-<J_{y}>J_{y}\right)  \label{exchange-term}%
\end{align}
$\mathcal{H}_{ex}^{\alpha_{1}}$ corresponds to the isotropic bilinear exchange
coupling while $\mathcal{H}_{ex}^{\alpha_{2}}$ corresponds to an anisotropic
bilinear coupling. The values of the isotropic ($n^{\alpha_{1}}$) and
anisotropic ($n^{\alpha_{2}}$) effective exchange coefficients\ can be
determined from the paramagnetic susceptibilities, measured with the field
along ($\chi_{\parallel c}$) and perpendicular\ ($\chi_{\bot c}$) to the
c-axis:%
\begin{align}
\chi_{\bot c} &  =\frac{C}{T-(n^{\alpha_{1}}-n^{\alpha_{2}})C}\nonumber\\
\chi_{\parallel c} &  =\frac{C}{T-(n^{\alpha_{1}}+2n^{\alpha_{2}}%
)C}\label{anistropic-sus}%
\end{align}
$C$ is the Curie constant. In Gd-based compounds, the two-ion
magnetoelasticity is related to the modification of the magnetic interactions
by the strains. For strains that preserve the initial tetragonal symmetry, the
two-ion magnetoelastic Hamiltonians can be written
as:\cite{Gignoux-Schmitt-1990}
\begin{align}
\mathcal{H}_{ME}^{\alpha_{1}} &  =-\left(  D_{\alpha_{1}}^{\alpha_{1}}%
\epsilon^{\alpha_{1}}+D_{\alpha_{1}}^{\alpha_{2}}\epsilon^{\alpha_{2}}\right)
<J>J,\nonumber\\
\mathcal{H}_{ME}^{\alpha_{2}} &  =-\left(  D_{\alpha_{2}}^{\alpha_{1}}%
\epsilon^{\alpha_{1}}+D_{\alpha_{2}}^{\alpha_{2}}\epsilon^{\alpha_{2}}\right)
\left(  2<J_{z}>J_{z}-<J_{x}>J_{x}-<J_{y}>J_{y}\right)  \label{two-ion-ME}%
\end{align}
where $D_{\alpha j}^{\alpha_{i}}$ are the two-ion magnetoelastic\ constants
and $\epsilon^{\alpha_{i}}$ are the normalized, symmetrized strains.
$\epsilon^{\alpha_{1}}$ is the volume strain while $\epsilon^{\alpha2}$ is the
axial strain acting on the c/a ratio:
\begin{align}
\epsilon^{\alpha_{1}} &  =\frac{1}{\sqrt{3}}(\epsilon_{xx}+\epsilon
_{yy}+\epsilon_{zz})\nonumber\\
\epsilon^{\alpha2} &  =\frac{1}{\sqrt{6}}(2\epsilon_{zz}-\epsilon
_{xx}-\epsilon_{yy})\label{sym-strains}%
\end{align}
\ 

The minimization of the free energy with respect to each strain yields the
equilibrium values of, say, the $\alpha$-strains:
\begin{align}
\epsilon^{\alpha_{1}}  &  =M_{\alpha_{1}}^{\alpha_{1}}<J>^{2}+M_{\alpha_{2}%
}^{\alpha_{1}}\left(  2<J_{z}>^{2}-<J_{x}>^{2}-<J_{y}>^{2}\right) \nonumber\\
\epsilon^{\alpha2}  &  =M_{\alpha_{1}}^{\alpha_{2}}<J>^{2}+M_{\alpha_{2}%
}^{\alpha_{2}}\left(  2<J_{z}>^{2}-<J_{x}>^{2}-<J_{y}>^{2}\right)
\label{strain-eqm-values}%
\end{align}
where $M_{\alpha j}^{\alpha_{i}}$ ($i,j$=1,2) is a combinations of the
$D_{\alpha j}^{\alpha_{i}}$ and the symmetrized elastic constant
$C_{0}^{\alpha_{k}}$($k$=1,2,12). The magnetoelastic couplings can be
experimentally probed by measuring the macroscopic length change under an
applied magnetic field. In a tetragonal symmetry, the relative length change,
$\lambda=\frac{\partial l}{l}$, measured along the ($\beta_{1},\beta_{2}%
,\beta_{3}$) direction when an external field is along the (x,y,z) direction
is given by:
\begin{equation}
^{\beta_{1}\beta_{2}\beta_{3}}\lambda_{xyz}=\frac{1}{\sqrt{3}}\epsilon
^{\alpha_{1}}+\frac{1}{\sqrt{6}}\epsilon^{\alpha_{2}}(2\beta_{3}^{2}-\beta
_{1}^{2}-\beta_{2}^{2})+\frac{1}{\sqrt{2}}\epsilon^{\gamma}(\beta_{1}%
^{2}-\beta_{2}^{2}) \label{length-variation}%
\end{equation}
where $\epsilon^{\gamma}$ corresponds to the strain mode that leads to a
symmetry lowering from tetragonal to orthorhombic. This $\epsilon^{\gamma}$
mode can be determined from the appropriate magnetostriction measurements,
namely:%
\begin{equation}
\epsilon^{\gamma}=\frac{1}{\sqrt{2}}(^{100}\lambda_{100}-^{010}\lambda_{100})
\label{epsilon-gamma}%
\end{equation}

The thermal and field variation of the measurable magnetostrictions arise from
the $T$- and $H$-dependence of $<J_{i}>$. The calculation of such a dependence
is usually carried out either by proper diagonalization of the total
Hamiltonian (mostly appropriate for the ordered state and higher applied
fields) or by perturbation methods (appropriate within the paramagnetic phase
and for small magnetic fields). Here we will be concerned only with the
paramagnetic case (for more details see Ref.\cite{Gignoux-Schmitt-1990}).
Utilizing the susceptibility formalism \cite{Gignoux-Schmitt-1990}, one gets
the following useful relations:
\begin{align}
\epsilon^{\alpha_{i}}  &  =\chi_{100}^{\alpha i}H^{2}\nonumber\\
\epsilon^{\alpha_{i}}  &  =\chi_{001}^{\alpha_{i}}H^{2} \label{strain-vs-H2}%
\end{align}
where $\chi_{j}^{\alpha_{i}}$ ($i$=1,2) is the strain field susceptibility
along the indicated direction. A complete description of the parastriction can
be obtained by combining these equations and Eqs.\ref{strain-eqm-values}
together with the relation:
\begin{equation}
<J_{i}>=\frac{\chi_{i}}{N_{A}g\mu_{B}}H;\ i=x,y,z;\text{ }N_{A}\text{ is the
Avogadro number} \label{paramagnetic-sus}%
\end{equation}

\section{Experimental}

Single crystals were grown by floating zone method.\cite{FZ-method} X-ray and
magnetic measurements revealed a single phase of an excellent quality. The
size and orientation of the rectangular spark-cut bars were chosen to suit the
requirement of the magnetization or the magnetostriction measurements. The
isofield and isothermal magnetization curves were measured by the extraction
method for temperatures down to 1.5 K and fields up to 160 kOe. The
sensitivity of the measurement is better than 5x10$^{-5}$ emu.

The magnetostriction measurement were performed using a high-accuracy
capacitance dilatometer which allows for measuring thermal expansion and
forced magnetostriction in the temperature range of 3 - 250 K and under
magnetic field up to 65 kOe. The typical resolution is better than 1
$\overset{\circ}{A}$. In addition, the set-up allows for the rotation of the
capacitance cell around the vertical axis of the cryostat while the field is
maintained in a horizontal direction. In this way, the angle between the
sample axis and the field direction can be varied across the range
$0-180^{\circ}$. The relative length variation for temperature scan at fixed
fields is defined as $\lambda_{H}=\left[  l(T)-l(T_{0})\right]  /l(T_{0})$
while that for field scan at fixed temperatures is defined as $\lambda
_{T}=\left[  l(H)-l(H=0)\right]  /l(H=0)$. Diamagnetic contributions and
demagnetizing fields are expected to be small and as such no correction for
their influences on the magnetization or magnetostriction curves were attempted.

\section{Results and Analysis}

Below we show representative magnetization and magnetostriction curves taken
within the paramagnetic and the ordered phase. By applying the field along the
(001) or (100) direction, it was possible to investigate separately the volume
distortion $\epsilon^{\alpha1}$, the axial distortion $\epsilon^{\alpha2}$ or
the orthorhombic distortion $\epsilon^{\gamma}$. The paramagnetic and
parastriction curves were analyzed according to susceptibility formalism given
in Sec.II. Based on the overall analysis of the ordered state, $H-T$ phase
diagrams for $H//\left(  100\right)  $ and $H//\left(  001\right)  $ are
constructed.%
\begin{figure}
[ptbh]
\begin{center}
\includegraphics[
height=8.8546cm,
width=12.6943cm
]%
{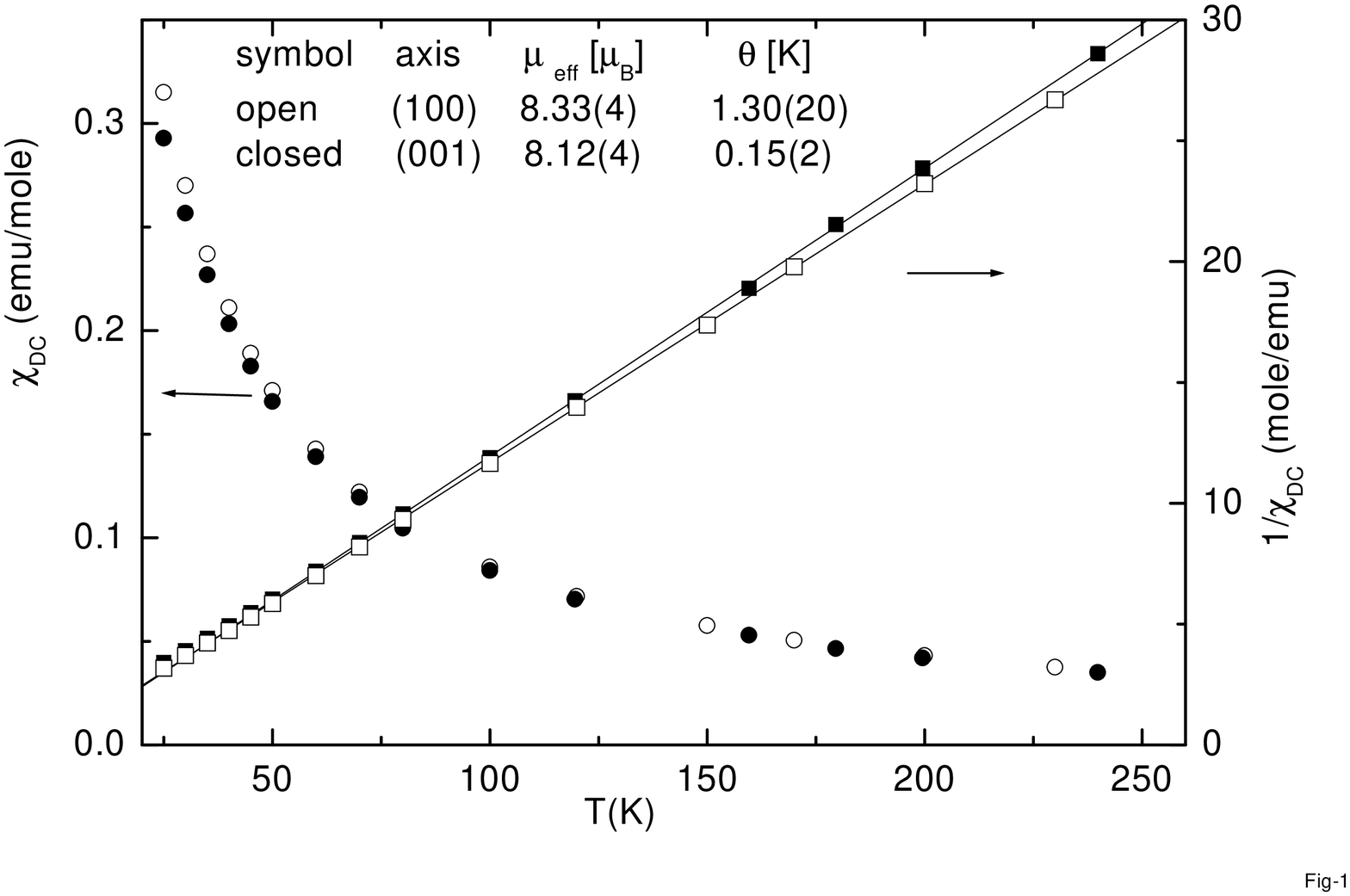}%
\caption{Thermal evolution of the paramagnetic susceptibilities (circles) and
their reciprocals (squares). The open and closed symbols represent,
respectively, measurements along the a-axis and the c-axis. Each point was
obtained from an Arrott plot ($M^{2}$ $versus$ $H/M$). The effective moment
and Curie-Weiss temperature for each orientation are given in the inset
table.}%
\end{center}
\end{figure}

\subsection{ Paramagnetic Phase}

\subsubsection{Magnetization}

The paramagnetic susceptibilities were deduced at each temperature (25
K$<T<$300 K) from the Arrott plot ($M^{2}$ versus $H/M$). The thermal
evolution of these susceptibilities along each of the (100) and (001)
direction (shown in Fig.1) reveals a weak anisotropic character, similar to
the one reported by Canfield et al \cite{Gd-single-crystal-Tdynamics}. From
the analysis of Fig.1 and according to Eq.\ref{anistropic-sus}, we obtained
$n^{\alpha_{1}}$=0.12($\pm$0.04) mole/emu and $n^{\alpha_{2}}$=-0.05($\pm
$0.04) mole/emu. The experimental error involved in the determination of these
$n^{\alpha_{1}}$ \ and $n^{\alpha_{2}}$ were dictated by the experimental
resolution, nonetheless, the determination of their signs and their magnitudes
are consistent with the results of Canfield et al
\cite{Gd-single-crystal-Tdynamics}: on the one hand, $n^{\alpha_{1}}$ is
positive and weak indicative of a small effective $\theta$. On the other hand,
the magnitude of $n^{\alpha_{2}}$ is twice smaller (indicative of a weak
paramagnetic anisotropy) and its sign is negative (i.e. $\chi_{\parallel
c}<\chi_{\bot c}$)\ implying that it is energetically favorable for the spins
to lie within the basal plane.

The effective moments deduced from the slope of the inverse susceptibilities
were found to be $\mu_{eff\left[  100\right]  }$=8.33(4) $\mu_{B}$ and
$\mu_{eff\left(  001\right)  }$=8.12(4)$\mu_{B}$; both are slightly higher
than the theoretical value ($\mu_{eff}$=7.94$\mu_{B}$). Such a small
discrepancy arises possibly due to matrix contribution: an
exchange-enhancement of the Ni-sublattice susceptibility and/or a polarization
of the conduction band.

\subsubsection{Parastriction}

Representative curves of the parastrictions $^{100}\lambda_{100}$,
$^{010}\lambda_{100}$, and $^{001}\lambda_{100}$ are shown in Fig.2. Similar
curves (not shown) were obtained as well for $^{i}\lambda_{010}\ $and
$^{i}\lambda_{001}\ (i=(100),(010),(001))$. In all cases, $\lambda$ is of the
order of few 10$^{-5}$, explaining the earlier report that magnetoelastic
effects were apparently absent in the temperature-dependent XRD
measurements.\cite{Gd-XRD-magnetoelastic}%

\begin{figure}
[ptbh]
\begin{center}
\includegraphics[
height=8.8546cm,
width=12.6943cm
]%
{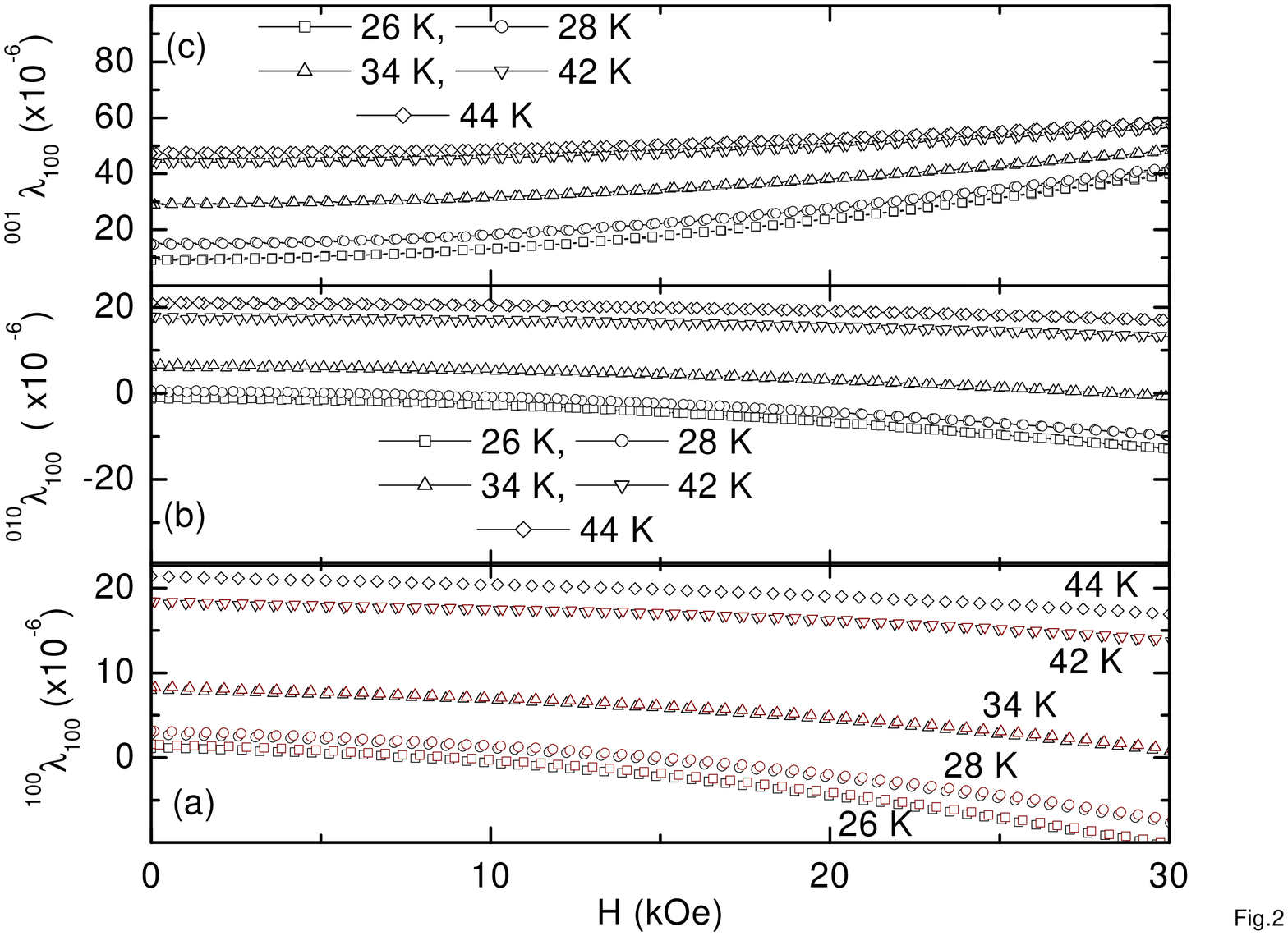}%
\caption{Representative parastriction isotherms with the field along the
a-axis. The relative length variation $\lambda$ is measured along (a) the
a-axis, (b) the b-axis, and (c) the c-axis.}%
\label{Fig.2}%
\end{center}
\end{figure}

Fig.3 shows the symmetrized, normalized\ strains $\epsilon^{\alpha1}$ and
$\epsilon^{\alpha2}$ which were constructed from parastrictions curves (as
those in Fig.2) according to Eq.\ref{length-variation}. Both $\epsilon
^{\alpha1}$ and $\epsilon^{\alpha2}$ are nonzero and follow faithfully the
quadratic field dependence (see Eq.\ref{strain-vs-H2}). These results coupled
with the tetragonal character of GdNi$_{2}$B$_{2}$C predict a magnetic
contribution to the thermal expansion, in addition to the conventional
contribution expected in the diamagnetic isomorph $R$Ni$_{2}$B$_{2}$C
($R$=Lu,Y, La).\ The symmetry lowering mode $\epsilon^{\gamma}$ (not shown) is
mostly field-independent ruling out any possibility for a field-induced
orthorhombic distortion within the paramagnetic phase.%

\begin{figure}
[ptbh]
\begin{center}
\includegraphics[
height=8.8546cm,
width=12.6943cm
]%
{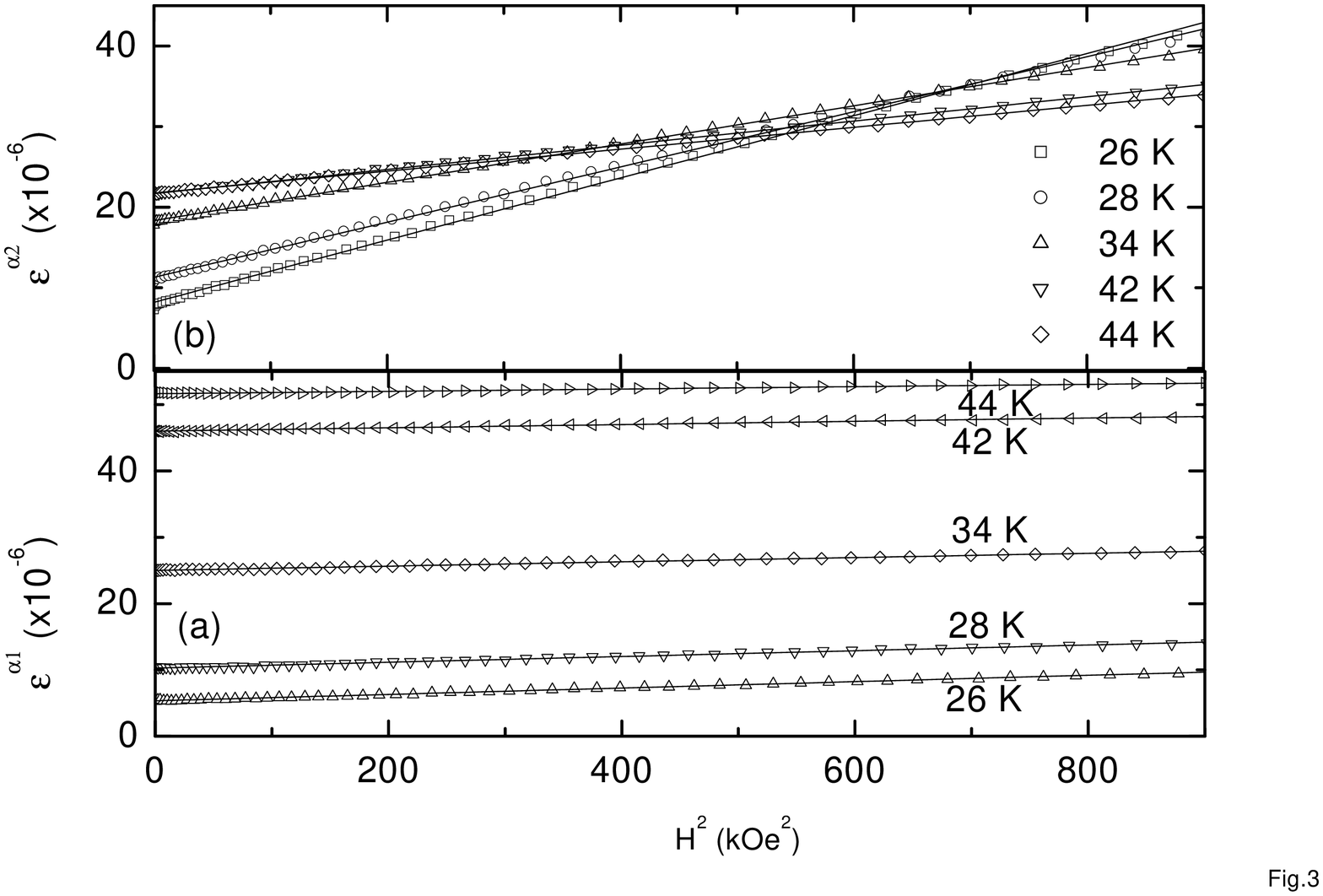}%
\caption{The normalized symmetrized strains (see Eq.\ref{sym-strains}) plotted
against the square of the field ($H//a$). (a) $\epsilon^{\alpha1}$ vs $H^{2}$
and (b) $\epsilon^{\alpha2}$ vs $H^{2}$. The continuous lines are the
least-square fit of Eq.\ref{strain-vs-H2} to the experimental points
(symbols).}%
\label{Fig.3}%
\end{center}
\end{figure}

Fig.4 shows the thermal evolution of $\chi_{100}^{\alpha_{1}}(T)$ and
$\chi_{001}^{\alpha_{1}}(T).$ These curves were obtained from the analysis of
Fig.3 according to Eq.\ref{strain-vs-H2}. Following the same procedure, we
obtained the\ $\chi_{100}^{\alpha_{2}}(T)$ and $\chi_{001}^{\alpha_{2}}(T)$
curves (shown in Fig.5). Within this paramagnetic range and according to
Eqs.\ref{length-variation} we get:-%

\begin{align}
\epsilon_{H//100}^{\alpha_{1}}  &  =\epsilon_{H//001}^{\alpha_{1}}\nonumber\\
\epsilon_{H//100}^{\alpha_{2}}  &  =\epsilon_{H//001}^{\alpha_{2}}
\label{length-paramagnet}%
\end{align}

then:%
\begin{align}
\chi_{100}^{\alpha_{1}}  &  =\chi_{001}^{\alpha_{1}}\nonumber\\
\chi_{100}^{\alpha_{2}}  &  =\chi_{001}^{\alpha_{2}}%
\end{align}

Within the experimental errors, these equalities were experimentally confirmed
in Figs.4-5: the analysis of the $\chi^{\alpha1}$ curves gave:\emph{ }%
\begin{align}
1/\sqrt{\chi_{100}^{\alpha_{1}}(T)\ }  &  =361(30)+41(1)\text{x}T\nonumber\\
1/\sqrt{\chi_{001}^{\alpha_{1}}(T)\ }  &  =324(73)+36(2)\text{x}T
\label{alhpa-(100)}%
\end{align}
while for the $\chi^{\alpha2}$ curves gave:\emph{ }%
\begin{align}
1/\sqrt{\chi_{100}^{\alpha_{2}}(T)\ }  &  =-27(17)+20(1)\text{x}T\nonumber\\
1/\sqrt{\chi_{001}^{\alpha_{2}}(T)\ }  &  =-37(26)+21(1)\text{x}T
\label{alpha-(001)}%
\end{align}
Thus, the thermal evolutions of the two $\alpha_{1}$- curves (as well the
$\alpha_{2}$-curves) are similar: the small numerical discrepancy is
attributed to artefacts stemming from the very small values of\ $\chi
^{\alpha_{i}}$ ($i=1,2$).%
\begin{figure}
[ptbh]
\begin{center}
\includegraphics[
height=8.8546cm,
width=12.6943cm
]%
{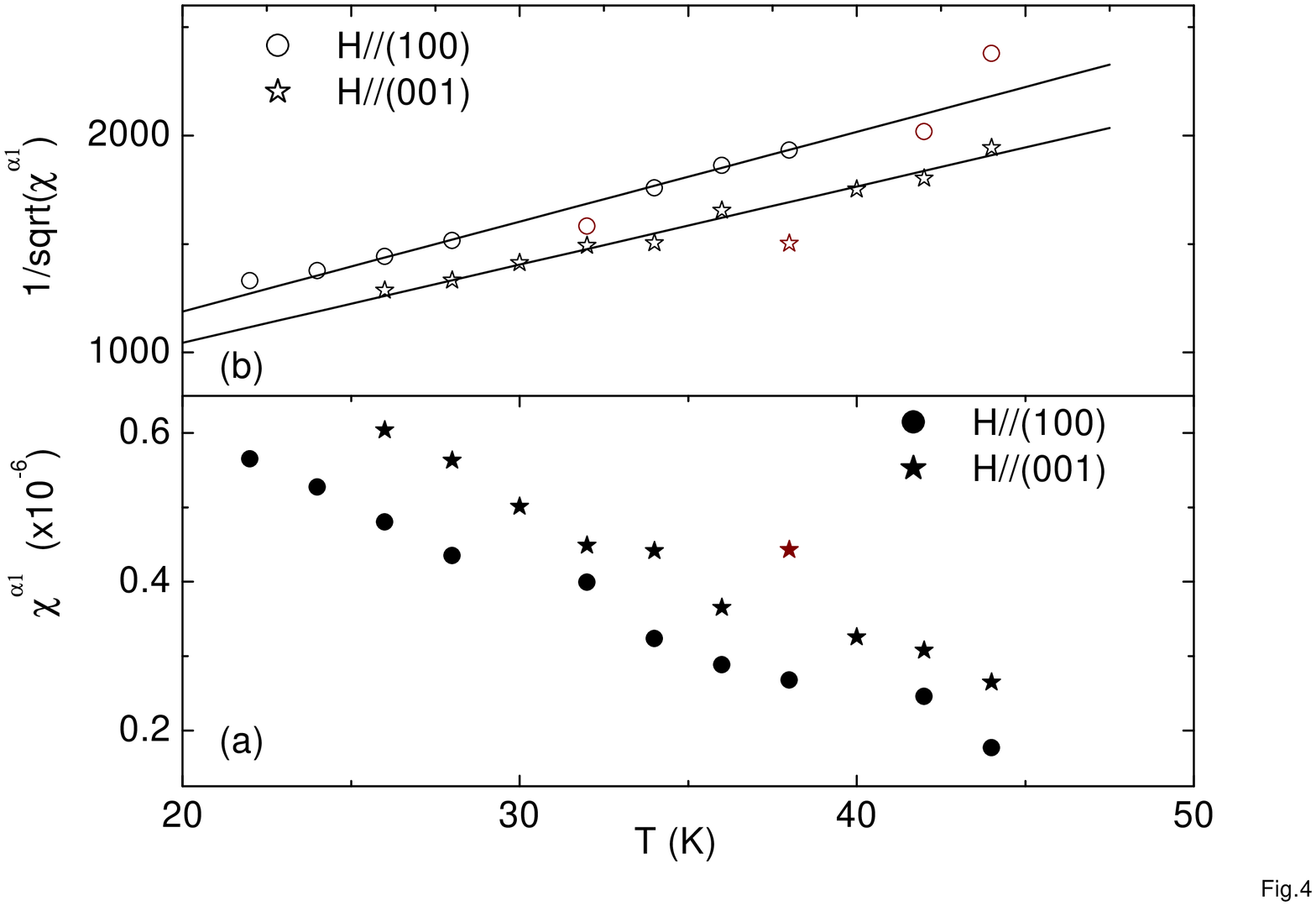}%
\caption{The thermal variation of the strain susceptibilities (a) $\chi
_{100}^{\alpha_{1}}(T)$, $\chi_{001}^{\alpha_{1}}(T)$ and (b) $\sqrt
{1/\chi_{100}^{\alpha_{1}}(T)\ }$ and $\sqrt{1/\chi_{001}^{\alpha_{1}}(T)\ }$.
The continuous lines are least-square fitting of the experimental points
(symbols).}%
\label{Fig.4}%
\end{center}
\end{figure}

\ From these expressions and the above-cited equations, $M_{\alpha j}%
^{\alpha_{i}}$ ($i,j$=1,2) can be straightforwardly calculated. However since
the values of the symmetrized elastic constant $C_{0}^{\alpha_{k}}$%
($k$=1,2,12)\ are unavailable, these calculation are not sufficient for
determining the extremely useful magnetoelastic constants $D_{\alpha
j}^{\alpha_{i}}$.%

\begin{figure}
[ptbh]
\begin{center}
\includegraphics[
height=8.8546cm,
width=12.6943cm
]%
{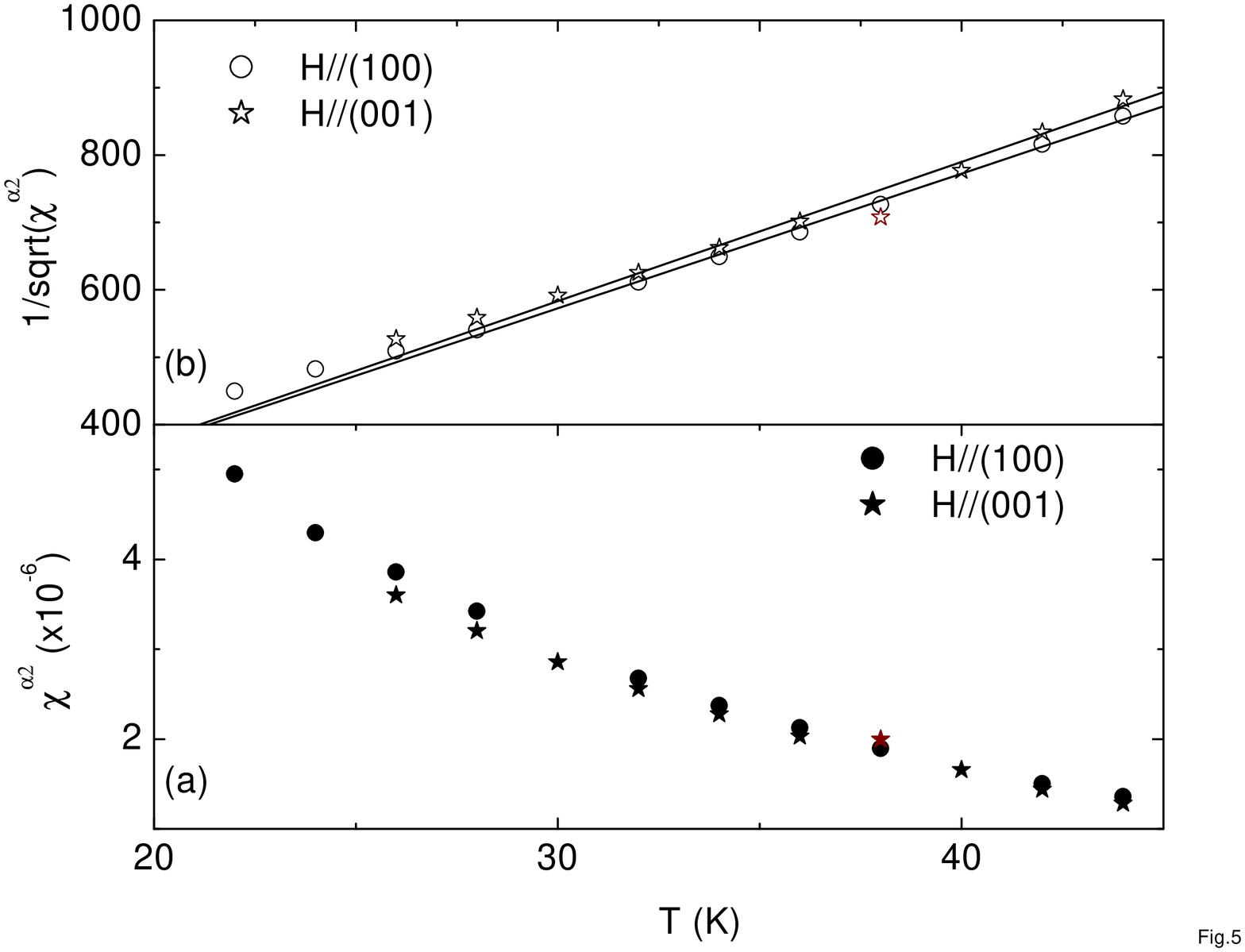}%
\caption{The thermal variation of the strain susceptibilities (a) $\chi
_{100}^{\alpha_{2}}(T),$ $\chi_{001}^{\alpha_{2}}(T)$ and (b) $\sqrt
{1/\chi_{100}^{\alpha_{2}}(T)\ }$ and $\sqrt{1/\chi_{001}^{\alpha_{2}}(T)\ }$.
The continuous lines are least-square fitting of the experimental points
(symbols).}%
\label{Fig.5}%
\end{center}
\end{figure}

\subsection{The Ordered Phase}

\subsubsection{Magnetization}

Fig.6a shows representative magnetization isotherms along the a-axis,
$M_{100}$. Each $M_{100}(H)$ isotherm increases monotonically with the field
and, for a certain field range, such an increase is linear. Moreover, for\ low
temperatures, $M_{100}(H)$ curves demonstrate a weak event at $H_{D}$
$\approx$12 kOe (defined as the field value at which $\left(  \partial
M/\partial H\right)  _{T}$ is a maximum) and, on a further field increase, a
saturation at a relatively high field, for instance $H_{S}$($1.5$ K$)=128$
kOe. It was observed that on increasing the temperature, the saturation
process becomes less pronounced. The thermal evolution of each of $H_{D}$ and
$H_{S}$ were plotted in Fig.11. The saturated moment (7.2$\mu_{B}$ at
$T=$1.5K) is weakly enhanced in comparison with the theoretically expected
value (7$\mu_{B}$). This enhancement is attributed to matrix contribution,
just as the above-mentioned enhancement of the effective moment.%
\begin{figure}
[ptbh]
\begin{center}
\includegraphics[
height=8.8568cm,
width=6.1637cm
]%
{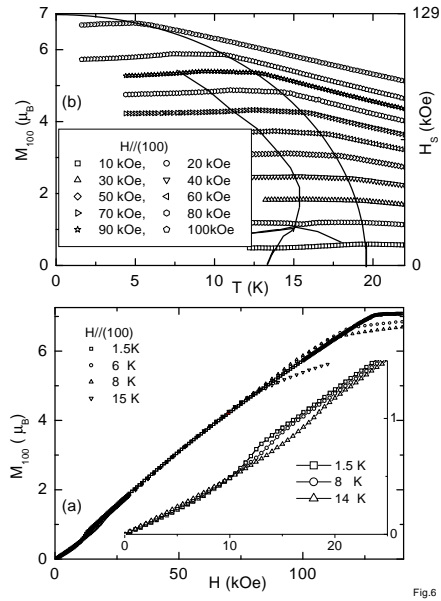}%
\caption{(a) Representative low-temperature magnetization isotherms along the
a-axis ($M_{100}$). The inset shows an expanded view of the low-field part of
the magnetization isotherm. (b) The thermal variation of the magnetization for
representative magnetic field ($H//a$). The continuous lines in Fig.6b
represent the thermal evolution of $H_{D}$, $H_{R},$ and $H_{S}$ as determined
from the analysis of Figs.6a,8-10b.}%
\label{Fig.6}%
\end{center}
\end{figure}

The thermal variation of the isofield $M_{100}(T)$ is shown in Fig.6b.
Characteristic features signaling magnetic events at $H_{D}$ and $H_{S}$ were
also observed and their values were found to agree with the values obtained
from the magnetization isotherms of Fig.6a. This agreement was convincingly
demonstrated by plotting in Fig.6b the thermal evolution of\ each of $H_{D}$
and $H_{S}$. Moreover, Fig.6b shows an extra curve, $H_{R}$, representing the
spin reorientation process (see below).

Representative magnetization isotherms along the c-axis ($M_{001}$) are shown
in Fig.7. These isotherms are very similar to the $M_{100}$ ones except that
(i) there are only two events namely at $H_{R}$ and $H_{S}$, (ii) the field
dependence of the magnetization at these events is more pronounced, and (iii)
$H_{R}$ was observed to have a different temperature dependence than the
corresponding one along the $a$-axis. However, for both orientation, the
saturation is reached at the same $H_{S}(T)$ curve. The thermal evolution of
$H_{S}$ and $H_{R}$ deduced from Fig.7 were plotted in Fig.12. It is worth
noting that within the experimental conditions of the extraction technique, no
hysteresis features or relaxation effects were observed.%
\begin{figure}
[ptbh]
\begin{center}
\includegraphics[
height=8.8568cm,
width=6.1637cm
]%
{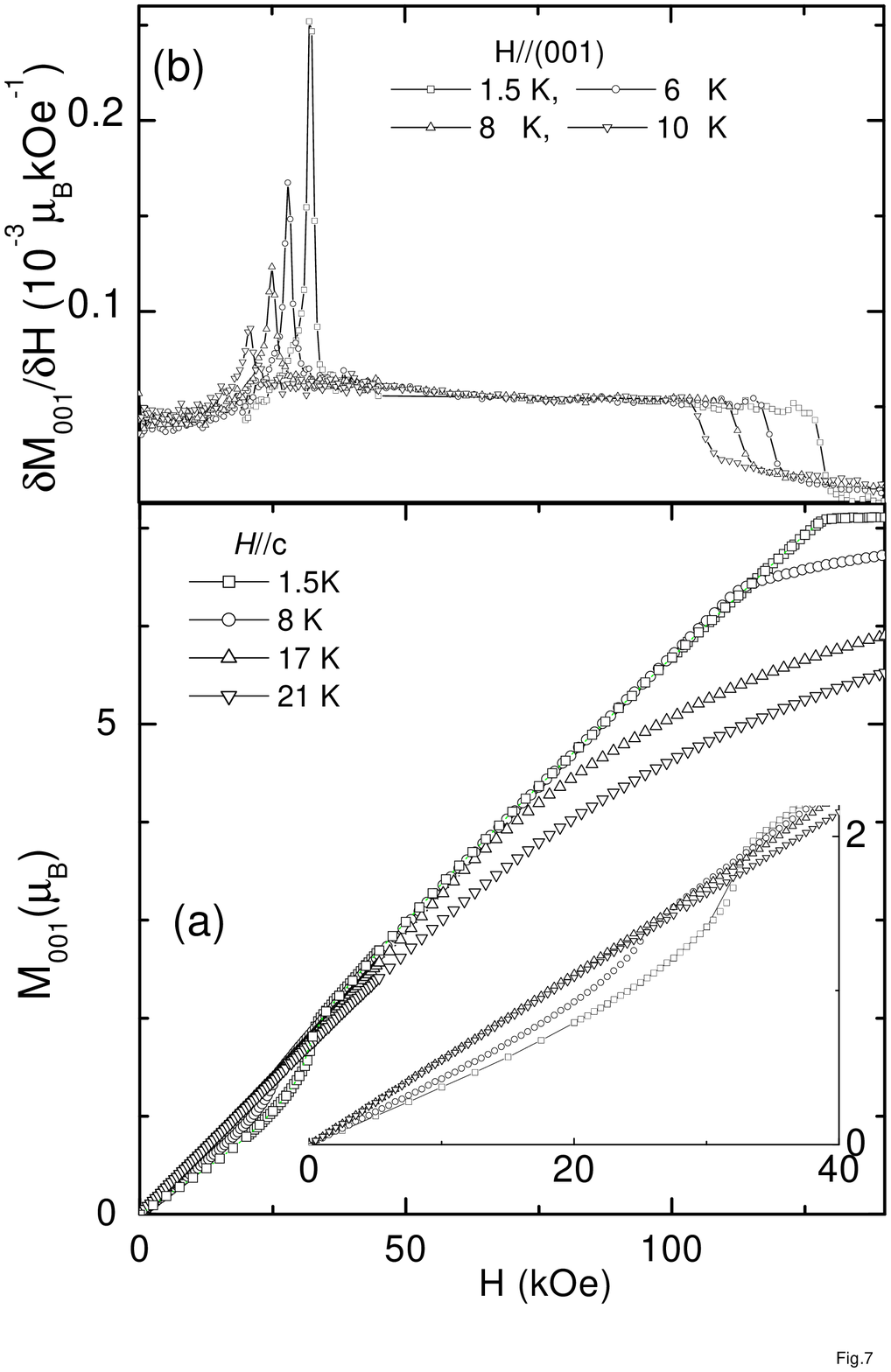}%
\caption{(a) Representative low-temperature magnetization isotherms along the
c-axis ($M_{001}$). The inset shows an expanded view of the low-field part of
selected magnetization isotherms. (b) Field derivatives of magnetization
isotherms showing the onset of $H_{R}(T)$ (the field at which the maximum is
attained) and $H_{S}(T)$ (the field of maximum inclination).}%
\label{Fig.7}%
\end{center}
\end{figure}

\subsubsection{Magnetostriction}

Fig.8a shows the forced magnetostriction curves, $^{100}\lambda_{100}(H)$, at
various temperatures. For lower temperatures, a field increase is accompanied
by a steep increase in $^{100}\lambda_{100}$, then a shallow maximum, and
afterwards a steady decrease with a weak slope. The initial steep increase is
very likely to be due to the domain-purifying influence of the applied field:
the field derivative of $\lambda$ shows a maximum at $H_{D}$ (compare Fig.6a
with Fig.8b). Surprisingly the $^{100}\lambda_{100}$ curves present a very
small hysteresis suggesting that the domain distribution has a weak dependence
on the sample history.%
\begin{figure}
[ptbh]
\begin{center}
\includegraphics[
height=8.8546cm,
width=12.6943cm
]%
{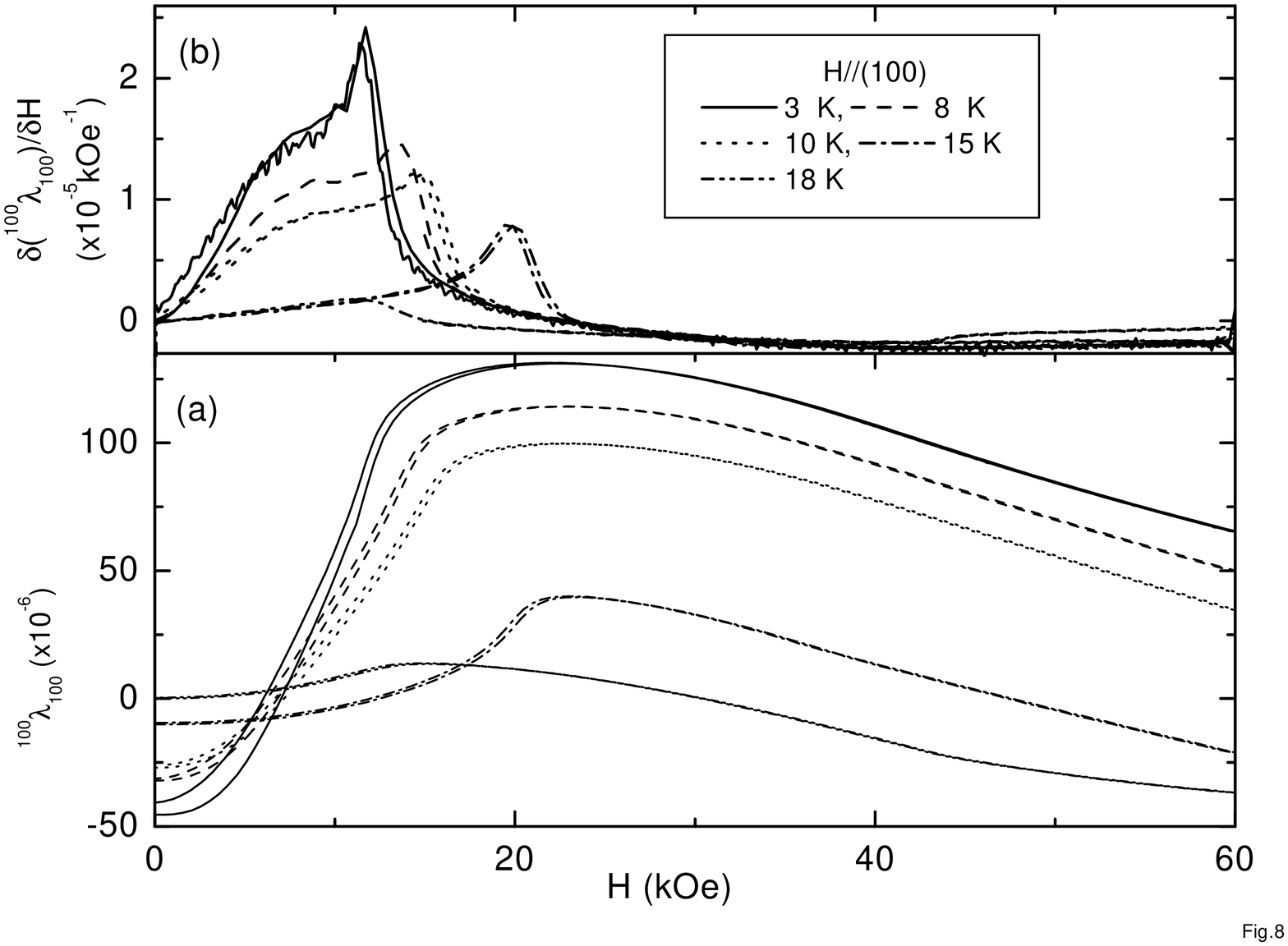}%
\caption{(a) Representative forced magnetostriction curves at various
temperatures. Measurements were taken while increasing and decreasing the
field. (b) Field derivative of these magnetostriction isotherms. Note that in
(b) the shoulder at the left hand side of $H_{D}(T)$ disappears when the
temperature is increased above $T_{R}(H=0)$ suggesting a correlation between
this shoulder and the reorientation process below $T_{R}(H)$.}%
\label{Fig.8}%
\end{center}
\end{figure}

The thermal evolution of the isofield $^{100}\lambda_{100}(T)$ (shown in
Fig.9) manifests magnetic events that are very much similar (though more
pronounced) to the ones observed in the magnetization (Figs.6b) and forced
magnetostriction curves (Fig8.a). The corresponding field and temperature
values at these events were plotted in Fig.11.

It is interesting to note that for intermediate fields, e.g. $H=15$ kOe in
Fig.9, on decreasing the temperature below $T_{N}$, $^{100}\lambda_{100}%
(T)$\ shows at first a slow increase till 17.4 K where a decrease commences
and continues till $T_{R}.$ Afterwards $^{100}\lambda_{100}(T)$ increases fast
but later tends to saturate to 1.3x10$^{-4}$ at $T=0$ K$.$ In contrast, for
lower (higher) fields, $^{100}\lambda_{100}(T)$ is monotonically decreasing
(increasing)\ with decreasing temperature, tending to saturate at $T$=0
K.\ For all applied fields, the signature of the event at $T_{R}(H)$ is
clearly observable but tends to fade away as higher fields are approached. It
is also interesting to compare the magnitude of this two-ion effect with the
one observed in typical single-ion compound such as ErNi$_{2}$B$_{2}$C
(Ref.\cite{Er-magnetostriction}): for zero field, the saturation value of
$^{100}\lambda_{100}$ in GdNi$_{2}$B$_{2}$C is $\sim$-0.5x10$^{-4}$ while in
ErNi$_{2}$B$_{2}$C is $\sim-$3x10$^{-4}$.%
\begin{figure}
[ptbh]
\begin{center}
\includegraphics[
height=8.8568cm,
width=6.1637cm
]%
{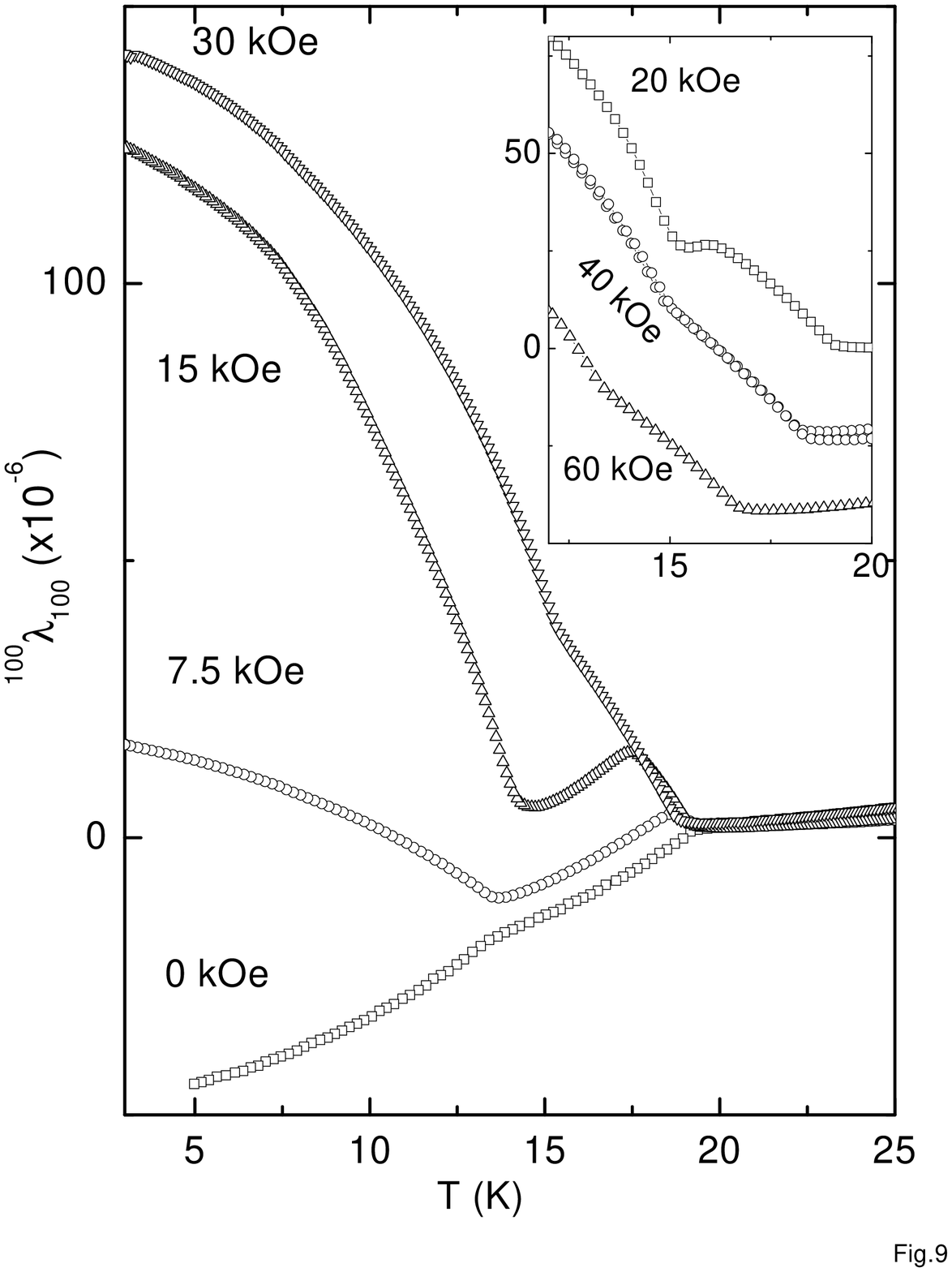}%
\caption{Representative thermal evolution of the magnetostriction for various
fields. Both the field and the length-measurement were oriented along the
a-axis. The inset shows, in an expanded scale, the thermal evolution of the
magnetostriction in the neighborhood of $T_{R}(H)$ and $T_{N}(H).$}%
\label{Fig.9}%
\end{center}
\end{figure}

Fig.10a shows an impressive demonstration of the magnetic domain effects
within the ordered state. In these curves, the angular dependence of various
$^{100}\lambda_{\theta}$ isotherms was monitored by rotating the single
crystal in a field of, say,\ 30 kOe. Furthermore, Fig.10b shows the
orthorhombic strain $\epsilon^{\gamma}$ for $H=$7.5, 30 kOe, constructed
according to Eq.\ref{epsilon-gamma}. Under $H$=30 kOe and a decreasing
temperature below $T_{N}$,\textit{ }$\epsilon^{\gamma}$ commences to increase
slowly till $T_{R}(=$15.5 K$)$ and afterwards with a faster rate till reaching
saturation of -2.6x10$^{-\text{ 4}}$. Across the borocarbides, this two-ion
orthorhombic distortion should vary roughly with the de Gennes factor and as
such for ErNi$_{2}$B$_{2}$C this partial contribution amounts to
$\epsilon^{\gamma}\approx-$0.4x10$^{-4}$ which is one third of the total
$\epsilon^{\gamma}$ $\approx-1.4$x10$^{-4}$ observed for ErNi$_{2}$B$_{2}$C at
2.5 K (Ref.\cite{Er-magnetostriction}): thus for borocarbides, the two-ion
magnetoelastic contribution (in particular $\epsilon^{\gamma}$) is not
negligible as compared to that of the single-ion effect.%
\begin{figure}
[ptbh]
\begin{center}
\includegraphics[
height=8.8568cm,
width=6.1637cm
]%
{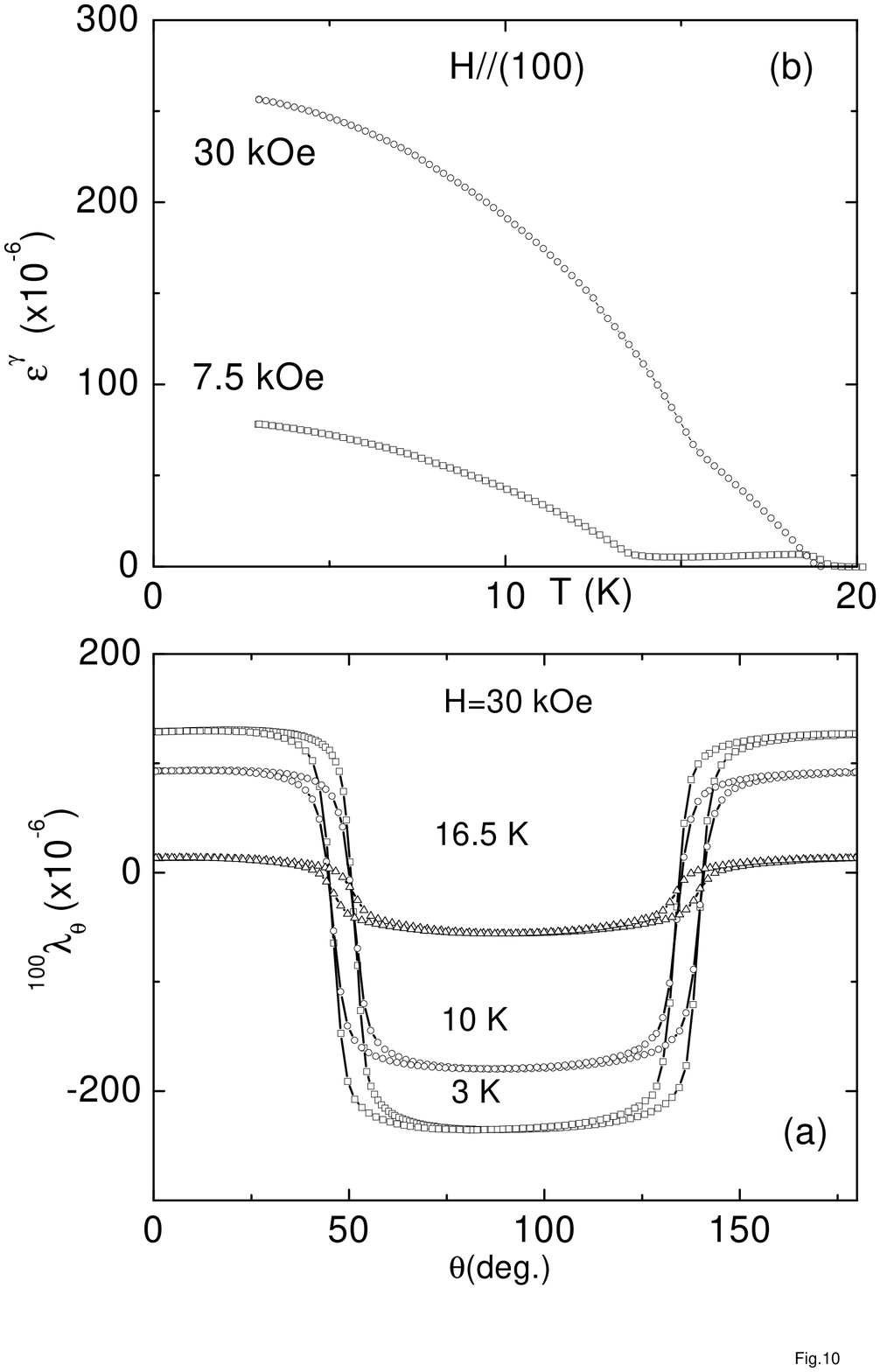}%
\caption{(a) Angular dependence of $^{100}\lambda_{\theta}$ isotherms under
$H=$ 30 kOe and for $T=$ 3 K, 10 K (below $T_{R}(H)$) and 16.5 K (above
$T_{R}(H)$). (b) the thermal evolution of $\epsilon^{\gamma}$ taken at $H=$7.5
kOe (below $H_{R}(T<13.5$ K$)$) and at $H=$30 kOe (above $H_{R}(T<13.5$
K$)$)$.$}%
\label{Fig.10}%
\end{center}
\end{figure}

\subsubsection{The $H-T\ $Phase Diagram}

Based on the analysis of the magnetization and magnetostriction curves, we
were able to construct the $H-T\ $phase diagrams with the field either along
the a-axis (shown in Fig.11) or along the c-axis (shown in Fig.12). In both
diagrams, the boundary describing the saturation process (when the exchange
field is completely counterbalanced) is well described by the relation
$H_{S}\left[  1-\left(  T/T_{N}\right)  ^{2}\right]  ^{\frac{1}{2}}$ kOe with
$H_{S}(T\rightarrow0)$=128.5(5) kOe and $T_{N}(H\rightarrow0)$=19.5 K. The
second boundary, denoted by $H_{R}(T)$, is attributed to the reorientation
process: its evolution for $H//c$ (much simpler than that for $H//a$) is given
by the relation $H_{R}$(1-$T/T_{R}$)$^{\frac{1}{3}}$ kOe where $H_{R}%
(T\rightarrow0)$=33.5(5) kOe and $T_{R}(H\rightarrow0)$=13.5 K. The
extrapolated values of the characteristic points$\ H_{S},$ $T_{N},$ $T_{R}$
are in excellent agreement with the reported
ones.\cite{Gd-XRES,Gd-single-crystal-Tdynamics,Gd-poly-high-field,Gd-chemical-composition,Gd-MES}
The third boundary $H_{D}(T),$ appearing only in the a-axis phase diagram,
(see Fig.11) is related to the domain effects as can be inferred from the
angular dependence of Fig.10.a. This $H_{D}(T)$ has the following features:
(i) the extrapolated $H_{D}(T\rightarrow0)$=12.0(5) kOe while $T_{D}%
(H\rightarrow0)$ $=T_{N}(H\rightarrow0)$ $=$19.5 K, (ii) it is almost constant
for $T<$ 8 K, increases slowly up to a maximum around 16 K and afterwards
decreases steadily till vanishes at $T_{N}$: such an increase of $H_{D}$ with
temperature is not usual for domain effects.%
\begin{figure}
[ptbh]
\begin{center}
\includegraphics[
height=8.8546cm,
width=12.6943cm
]%
{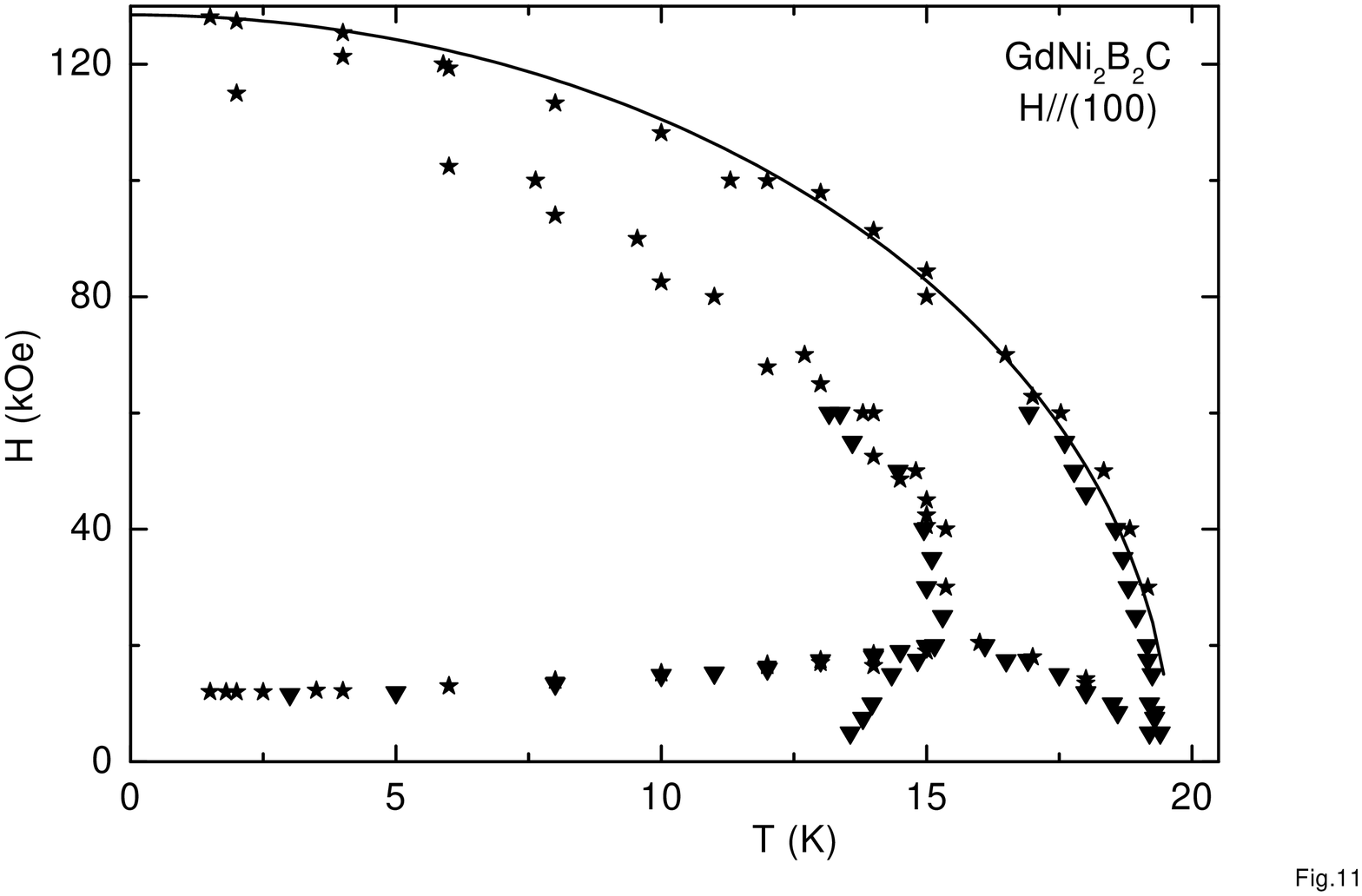}%
\caption{The $H-T$ phase diagram along the a-axis as compiled from the
magnetization ($\bigstar$) and magnetostriction ($\blacktriangledown$)
measurements. The continuous line represents the best fit of the experimental
points to $H_{S}$(1-$\left(  T/T_{N}\right)  ^{2}$)$^{\frac{1}{2}}$ kOe where
$H_{S}$ and $T_{N}$ were found, respectively, to be 128.5(5) kOe and 19.5 K.}%
\label{Fig.11}%
\end{center}
\end{figure}

\section{Discussion}

The modulated state in GdNi$_{2}$B$_{2}$C (and borocarbides in general) is
usually related to the nesting features in their electronics
structures.\cite{Generalized-X} Below, we discuss the stability of this state
against field and temperature variation. First MFA arguments are applied to
account for $H_{S}$, $H_{D}$, and the linearity of the magnetization
isotherms. Next, anisotropic perturbation are introduced and and afterwards
the anisotropic exchange interaction and magnetoelastic couplings are
considered.%
\begin{figure}
[ptbh]
\begin{center}
\includegraphics[
height=8.8546cm,
width=12.6943cm
]%
{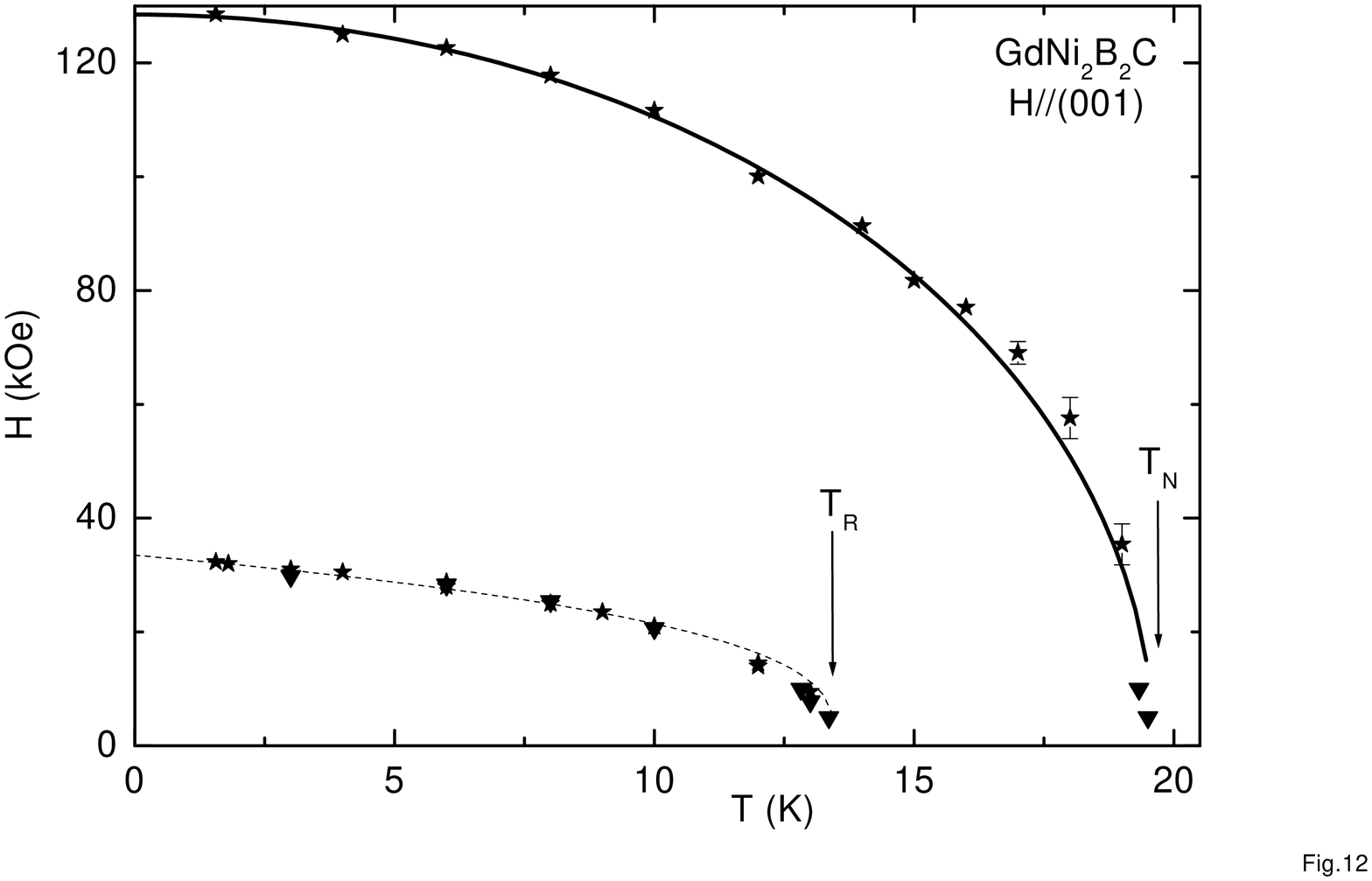}%
\caption{The $H-T$ phase diagram ($H//(001)$) as compiled from the
magnetization ($\bigstar$) and magnetostriction ($\blacktriangledown$) curves.
The continuous line represent the best fit of the data to $H_{S}\left[
1-\left(  T/T_{N}\right)  ^{2}\right]  ^{\frac{1}{2}}$ kOe with $H_{S}%
(T\rightarrow0)$=128.5(5) kOe and $T_{N}(H\rightarrow0)$=19.5 K. The dashed
lines represent the best fit to $H_{R}$(1-$T/T_{R}$)$^{\frac{1}{3}}$ kOe where
$H_{R}(T\rightarrow0)$=33.5(5) kOe and $T_{R}(H\rightarrow0)$=13.5 K.}%
\label{12}%
\end{center}
\end{figure}

Among the borocarbides, GdNi$_{2}$B$_{2}$C has the strongest exchange
interactions and the weakest magnetic anisotropy. The strength of the exchange
interactions (denoted as $H_{ex}$, or $J_{eff}$) is reflected as a high value
of $T_{N}$ and $H_{S}$ (see Figs.11-12) while the weakness of the magnetic
anisotropy (denoted as $H_{an}$) is manifested as low $H_{D}$ and as a weak
difference between the parallel and perpendicular paramagnetic susceptibility.
MFA arguments can be used to calculate an effective value for each of $H_{ex}%
$, $J_{eff}$, and $H_{an}$ from the observed values of $T_{N}$, $H_{S},$ and
$H_{D}$. Let us, at first, ignore the anisotropic feature ($H_{an}$=0). Then,
from the\ relation $zJ_{eff}=\frac{3T_{N}}{2S(S+1)}$ (z is an effective
nearest neighbors), $zJ_{eff}$ was calculated $\approx$1.86 K. Substitution of
this value into the relation $H_{S}=2zJ_{eff}S(S+1)/g\mu_{B}S$ gave $H_{S}$
$\approx$125 kOe, a reasonable value for the effective saturation field. Next,
taking into account the case $H_{an}\neq0$ and using Eqs.\ref{MF-Hs-Ha}, one
is able to calculate an effective value for each of $H_{ex}$ and $H_{an}$ and
furthermore to account for the observed linearity of $M_{100}(H_{D}<H<H_{S})$
: \cite{Smart}%
\begin{align}
H_{S} &  =(H_{ex}-H_{an})\nonumber\\
H_{D} &  =\sqrt{H_{an}H_{ex}-H_{an}^{2}}\nonumber\\
\left\langle M_{100}\right\rangle /M_{s} &  =H/H_{S},\text{ for }H_{D}%
<H<H_{S}\label{MF-Hs-Ha}%
\end{align}
Substituting the extrapolated $H_{S}(T\rightarrow0$ K) and $H_{D}%
(T\rightarrow0$ K) (see Fig.11 below) one obtains the satisfactory value
$H_{ex}(T\rightarrow0$ K) = 129.6(5) kOe and $H_{an}(T\rightarrow0$ K)=1.1(3)
kOe: as expected for a Gd-based compound, the former is much stronger than the latter.

In the above treatment, the zero-temperature magnetic structure is implicitly
assumed as a collinear 3d AF state. Rather, for $T_{R}(H=0)<T<T_{N},$ the
magnetic structure is a modulated state with no c-component and the ordered
moments are distributed among the a- or b-polarized domains. Then applying a
field along the a-axis would force the a-polarized domains to align
perpendicular to the field at $H_{D}(T>T_{R})$ and later on to achieve
saturation at $H_{S}(T>T_{R})$. For the same temperature range, a field
applied along the c-axis is already perpendicular to the domains and as such
only the saturation transition would appear. On the other hand for
$T<T_{R}(H=0),$ the spontaneous emergence of the c-component at $H_{R}(T)$
constitutes an extra event for both field orientations: thus for
$H//$a-axis,\ there are three transitions (at $H_{D}$, $H_{R}$, and $H_{S})$
while for the $H//$c-axis, there are only two transitions (at $H_{R}$ and
$H_{S})$.

Within the ordered phase, the manifestation of the magnetoelastic coupling and
anisotropic exchange interactions is much more pronounced, leading to a
substantial perturbation of the nuclear and the magnetic structures. To
demonstrate the importance of these perturbations, let us divide the ordered
phase into two regions depending on the magnitude of $\epsilon^{\gamma}$. The
first region is$\ T_{R}<T<T_{N}$ wherein the low-field $\epsilon^{\gamma}$ is
relatively weak (see Fig.10) and as such is hardly sufficient to modify
(substantially)\ the nuclear structure or the modulated magnetic state. The
second region is $T<T_{R}$ wherein the magnitude of the low-field
$\epsilon^{\gamma}$ is very large (see Fig.10) and increases whenever the
temperature is decreased or the field is increased. As a consequence, the
nuclear structure is distorted ($\frac{a}{b}<1$ and the four-fold symmetry of
the basal plane is reduced to a two-fold symmetry). In addition, there is an
accompanying dramatic modification of the magnetic structure: as the
temperature is decreased, the anisotropic character of the exchange
interactions (just as the magnitude of $\epsilon^{\gamma}$) is enhanced and
the magnetic structure is more and more perturbed towards a squared-up state
wherein the magnetic moments approach equal amplitudes and progressively bunch
towards an orientation within the b-c plane. While the square-up character can
be understood based on entropy arguments, the bunching of the moments is most
probably induced by the above mentioned perturbations.

By demonstrating the presence of magnetoelastic and anisotropic exchange
couplings in the ordered phase of GdNi$_{2}$B$_{2}$C, this works confirms the
far-sight analysis of Detlefs et al \cite{Gd-XRES} and Tomala et al
\cite{Gd-MES}. On the one hand, Detlefs et al attributed the strong asymmetry
in the line shape of the resonant XRES below $T_{R}(H=0)$ as being due to a
straining. Consequently, their XRES spectra were analyzed as being composed of
two peaks (thus the asymmetric character) below $T_{R}(H=0)$ and as a single
line above $T_{R}(H=0)$. It is assuring to note that the thermal evolution of
the position and intensity of these XRES peaks correlate very well with
$\epsilon^{\gamma}$\ (compare Fig.7 of Ref.\cite{Gd-XRES} with our Fig.10).
Tomala et al, on the other hand, analyzed their $^{155}$Gd Mossbauer spectra
assuming a transverse (moments along b-axis) sine-modulated state above
$T_{R}(H=0)$ and a strongly \textit{squared-up} and \textit{bunched} state
below $T_{R}(H=0)$\textit{.}

\section{Conclusion}

We presented experimental results on single crystal of GdNi$_{2}$B$_{2}$C. The
parastriction and paramagnetic features, well accounted for within the
susceptibility formalism, demonstrated the presence of weak anisotropic
exchange couplings and two-ion $\epsilon^{\alpha_{1}}$ and $\epsilon
^{\alpha_{2}}$ strain modes. The strength of the latter coefficients indicate
that the strain derivative of the coupling constant at the Brillouin zone
centre have significant amplitudes. The ordered state, on the other hand, is
characterized by very rich field and temperature phase diagrams that manifest
high values of $T_{N}$ and $H_{S}$ (indicative of strong exchange
interactions), a spin reorientation process$\ $at $H_{R}(T)$, and a
domain-purifying field $H_{D}$. Anisotropic exchange interactions and
relatively strong magnetoelastic couplings (particularly the orthorhombic
distortion mode) were observed, lending support to the reported modification
of the magnetic structure below $T_{R}$.

It is worth recalling that the modulated state of GdNi$_{2}$B$_{2}$C within
the temperature range $T_{R}<T<T_{N}$ is very much similar to that of
ErNi$_{2}$B$_{2}$C within the range 2 K$<T<T_{N}$, even though the dominant
aligning forces operating in each isomorphs are very much different. Since the
exchange interaction and the magnetoelastic couplings are driven by the
indirect RKKY couplings and that de Gennes scaling holds well among the heavy
borocarbides,\cite{Canfield-physics-today} then both the anisotropy exchange
interactions and magnetoelastic couplings must be present (though with varying
strength depending on de Gennes factor) in other borocarbide magnets. As such,
these perturbations must be explicitly considered when discussing the magnetic
properties (in particular the $H-T\ $phase diagrams) of these magnets.

\begin{acknowledgement}
M. ElM. acknowledge partial financial support from the Japan Society for the
promotion of science, the University of Joseph Fourier \ of Grenoble, and the
Brazilian agencies CNPq and FAPERJ.
\end{acknowledgement}


\begin{thebibliography}{99}                                                                                               %


\bibitem {RNi2B2C-mag-xrd}J.W. Lynn, S. Skanthakumar, Q. Huang, S. K. Sinha,
Z. Hossain, L.C. Gupta, R. Nagarajan, C. Godart, Phys. Rev. B \textbf{55},
6584 (1997).

\bibitem {Er-HT-diagram}P. C. Canfield, S. L. Bud'ko, and B. K. Cho, Physica C
\textbf{262}, 249 (1996); S. L. Bud'ko and P. C. Canfield, Phys.\ Rev. B
\textbf{61}, R14 932 (2000); A. J. Campbell, D. McK. Paul and G. J. MacIntyre,
Solid State Com \textbf{115}, 213 (2000).

\bibitem {Ho-HT-Diagram}C. Detlefs, F. Bourdarot, P. Burlet, P. Dervenagas, S.
L. Bud'ko and P. C. Canfield, Phys.\ Rev. B \textbf{61}, R14916 (2000); P. C.
Canfield, S. L. Bud'ko, B. K. Cho, A. Lacerda, D. Farrell, E.
Johnston-Halperin, V. A. Kalatsky, and V. L. Pokrovsky, Phys.\ Rev. B
\textbf{55}, 970 (1997); A. J. Campbell, D. McK.Paul and G. J. McIntyre,
Phys.\ Rev. B \textbf{61}, 5872 (2000).

\bibitem {Dy-HT-diagram}P.C. Canfiedl and S.L. Bud'ko, J. Alloys Compd.
\textbf{262-263}, 169 (1997).

\bibitem {Dy-structure}P. Dervenagas, J. Zarestky, C. Stassis, A. I. Goldman,
P. C. Canfield, and B. K. Cho, Physica B \textbf{212}, 1 (1995).

\bibitem {Tb-structure-anistropy-WeakFM}P. Dervenagas, J. Zarestky, C.
Stassis, A. I. Goldman, P. C. Canfield, and B. K. Cho, Phys.\ Rev. B
\textbf{53}, 8506 (1996); B. K. Cho, P. C. Canfield, and D. C. Johnston,
Phys.\ Rev. B \textbf{53}, 8499 (1996).

\bibitem {Er-Mag-structure}J. Zarestky, C. Stassis, A. I. Goldman, P. C.
Canfield, P. Dervenagas, B. K. Cho, and D. C. Johnston, Phys.\ Rev. B
\textbf{51}, 678 (1995); S. K. Sinha, J. W. Lynn, T. E. Grigereit, Z. Hossain,
L. C. Gupta, R. Nagarajan, and C. Godart, Phys. Rev. B \textbf{51}, 681 (1995).

\bibitem {Gd-XRES}C. Detlefs, A. I. Goldman, C. Stassis, P. C. Canfield, B. K.
Cho, J.P. Hill, and D. Gibbs, Phys.\ Rev. B \textbf{53}, 6355 (1996).

\bibitem {Theory-exchange-CEF-Phase-diagram}A. Amici and P. Thalmeier,
Phys.\ Rev. B \textbf{57}, 10684 (1998); V. A. Kalatsky and V. L. Pokrovsky,
Phys.\ Rev. B \textbf{57}, 5485 (1998).

\bibitem {Er-magnetostriction}M. Doerr, M. Rotter, M. El Massalami, S.
Sinning, H. Takeya and M. Loewenhaupt, J. Phys: Cond. Matter \textbf{14} 5609 (2002)..

\bibitem {Ho-distortion}A. Kreyssig, M. Loewenhaupt, J. Freudenberger, K.-H.
Muller, and C. Ritter, J. Appl. Phys, \textbf{85}, 6058 (1999); G. Oomi, T.
Kagayama, H. Mitamura, T. Goto, B. K. Cho, P. C. Canfield, Physica B 294-295,
229 (2001).

\bibitem {Dy-magnetostriction}C. Sierks, M. Doerr, A. Kreyssig, M.
Loewenhaupt,Z. Q. Peng, K. Winzer, J. Mag. Mag. Mat. \textbf{192}, 473 (1999).

\bibitem {Tb-dichroism-HRMXRD}C. Song, J. C. Lang, C. Detlefs, A. Letoublon,
W. Good, J. Kim, D. Wermeille, S. L. Budko, P. C. Canfield, A. I. Goldman,
Phys. Rev. B \textbf{64}, 020403 (2001); C. Song, D. Wermeille, A. I. Goldman,
P. C. Canfield, J. Y. Rhee and B. N. Harmon, Phys. Rev B \textbf{63}, 104507 (2001).

\bibitem {Gd-single-crystal-Tdynamics}P. C. Canfield, B. K. Cho, K. W. Dennis,
Physica B \textbf{215}, 337 (1995).

\bibitem {R1221-structure}T. Siegrist, H.W. Zandbergen, R.J. Cava, J.J.
Krajewski and W. F.Peck, Nature \textbf{367}, 254 (1994); T. Siegrist, R.J.
Cava, J.J. Krajewski and W. F.Peck, Jr, J. Alloys Compd. \textbf{216}, 135 (1994).

\bibitem {Gd-poly-high-field}M. El Massalami, B Giordanengo, J. Mondragon, E.
M. Baggio-Saitovitch, A. Takeuchi, J. Voirn and A. Sulpice, J. Phys:Condens
Matter \textbf{50}, 10015 (1995); M. El Massalami, S.L. Bud'ko, B.
Giordanengo, M.B. Fontes, J.C. Mondragon and E. M. Baggio-Saitovitch, Phys.
Status Solidi B \textbf{5}, 489 (1995).

\bibitem {Gd-chemical-composition}C. Godart, I. Felner, H. Michor, G.
Hilscher, E. Tominez, E. Alleno, J. Alloys Compd. \textbf{277}, 642 (1998).

\bibitem {Gd-MES}K. Tomala, J. P. Sanchez, P. Vulliet, P. C. Canfield, Z.
Drzazga, A. Winiarska,Phys. Rev. B\textbf{\ 58}, 8534 (1998); D. R. Sanchez,
H. Micklitz, M. B. Fontes, et al., J. Phys: Condens Matter,\textbf{\ 9},
L299-L302 (1997); F.M. Mulder, J. V. V. J. Brabers, R. Cochoorn, R.C. Thiel,
K. H. J. Buschow, F. R.de Boer, J. Alloys Compd. \textbf{217}, 118 (1995).

\bibitem {Gignoux-Schmitt-1990}P. Morin, J. Rouchy, and D. Schmitt, Phys. Rev.
B \textbf{37}, 5401 (1988).

\bibitem {FZ-method}H. Takeya, T. Hirano, K. Kadowaki, Physica C \textbf{256},
220 (1996).

\bibitem {Gd-XRD-magnetoelastic}A. Lindbaum and M. Rotter, preprint submitted
to Elsevier (2002).

\bibitem {Generalized-X}J.Y. Rhee, X. Wang, and B. N. Harmon, Phys. Rev.B
\textbf{51, }15 585 (1995).

\bibitem {Smart}J.S. Smart, \textit{Effective Field Theories of Magnetism},
(W.B.Saunders Company,1966) p.101; J.S. Smart, in \textquotedblright%
\textit{Magnetism.III}\textquotedblright, ed. G.Rado and H. Suhl, (Academic
Press, London, N.Y., 1963) p.63.

\bibitem {Canfield-physics-today}P. C. Canfield, O. L. Gammel, and D. J.
Bishop, Phys. Today, \textbf{51}, 40 (1998).
\end{thebibliography}
\end{document}